\documentclass[%
prl,%
twocolumn,%
]{revtex4}%

\usepackage{graphicx}

\begin{document}

\title{Comment on {\em Sixth-Order Vacuum-Polarization Contribution to the Lamb Shift of Muonic Hydrogen\/}
by T.~Kinoshita, and M.~Nio, Phys.~Rev.~Lett. {\bf 82}, 3240
(1999)}
\author{Vladimir G. Ivanov}
%\email{ivanov.vg@gao.spb.ru}
\affiliation{Pulkovo Observatory, 196140, St.Petersburg, Russia}
\affiliation{D. I. Mendeleev Institute for Metrology (VNIIM), St. Petersburg 198005, Russia}
\author{Evgeny Yu. Korzinin}
\affiliation{D. I. Mendeleev Institute for Metrology (VNIIM), St. Petersburg 198005, Russia}
\author{Savely G. Karshenboim}
\email{savely.karshenboim@mpq.mpg.de} \affiliation{D. I. Mendeleev
Institute for Metrology (VNIIM), St. Petersburg 198005, Russia}
\affiliation{Max-Planck-Institut f\"ur Quantenoptik, 85748 Garching,
Germany}

\maketitle

Recently, while performing a calculation of the $\alpha^2$
corrections to HFS interval in muonic hydrogen \cite{russian}, we
had to calculate contributions of the third order {\em bound-state}
perturbation theory (PT). The general expression for those
corrections is of the form (see, e.g., \cite{III,Bethe})
\begin{equation}\label{e3}
%\[
\Delta E^{(3)}(ns)=
%\]
%\begin{equation}\label{e3}
\langle \Psi_{ns}|\delta
 V\widetilde{G} \Bigl[\delta V-\Delta E_{ns}^{(1)}\Bigr]\widetilde{G}
 \delta V|\Psi_{ns}\rangle
 \;,
\end{equation}
where $\Delta E_{ns}^{(1)}=\langle\Psi_{ns}|\delta
V|\Psi_{ns}\rangle$, $\delta V$ is a sum of all perturbations under
consideration and $\Psi_{ns}
%({\bf r})
$ and $\widetilde{G}
%({\bf r}',{\bf r''})
$ are the wave function and the reduced Green function,
respectively, of the unperturbed problem (i.e., of the
non-relativistic Coulomb problem in our case).

In contrast to the scattering PT the bound-state PT contains certain
subtractions.
% A well known example of such a subtraction for
% the second-order terms is the fact that the scattering PT contains a
% complete Green function, while the bound-state PT deals with the
% reduced Green function.
In particular, in the third order, the bound-state PT (see
Eq.~(\ref{e3})) in addition to an ordinary contribution ($\sim
\delta V$) involves one more term ($\sim \Delta E_{ns}^{(1)}$). We
refer to the former as to a `main' term and to the latter as to a
`subtraction' term.

\begin{figure}[ptbh]
 \begin{center}
 \includegraphics[height=2cm]{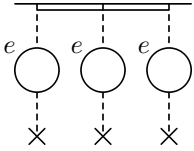}
 \end{center}
 \caption{The
$\alpha^5 m$ correction to the Lamb shift in muonic hydrogen: the
only contribution of the third order of non-relativistic
perturbation theory (cf. Fig.~5{\em c} in \protect\cite{kinoshita})
 \label{f:Lamb3}}
 \end{figure}

For muonic hydrogen the $\alpha^2$ contributions into the HFS
interval are similar to one of the $\alpha^3$ corrections to the
Lamb shift (see Fig.~\ref{f:Lamb3}), which was previously calculated
for the $2s$ and $2p$ states in \cite{kinoshita} (see the third line
in Eq.~(25) in \cite{kinoshita}),
\begin{equation}\label{kn}
\Delta E(2p-2s)=0.002535(1)\frac{\alpha^5}{\pi^3}
 m_r c^2\;,
\end{equation}
where $m_r$ is the reduced mass for muonic hydrogen and $\alpha$
stands for the fine structure constant.

To cross-check our calculations on HFS \cite{russian}, we have also
calculated contribution of diagram in Fig.~\ref{f:Lamb3} into the
$2s-2p$ splitting and found
\begin{eqnarray}
\Delta E(2p-2s) &=&
\Biggl[\bigl(-7.3861\cdot10^{-6}+0.3511\cdot10^{-6}\bigr)\nonumber\\
&&[ - \bigl(-0.002\,5412+0.001\,3661\bigr)\Biggr]
\frac{\alpha^5}{\pi^3} m_r c^2 \nonumber\\
\label{ours} &=& 0.0011681\frac{\alpha^5}{\pi^3} m_r c^2\;,
\end{eqnarray}
where the first parentheses are for the $2p$ contribution, while the
second are for the $2s$ one; each consists of the main and the
subtraction term as introduced in (\ref{e3}).

The results in (\ref{kn}) and (\ref{ours}) disagree. Our result
confirms calculations \cite{kinoshita} for the main terms for the
$2s$ and $2p$ states, while the difference originates from the fact
that the subtraction terms are missing in \cite{kinoshita}, as we
later learned \cite{kinoshitaplus} from the authors of the paper.

The work was in part supported by DFG (under grant \# GZ 436 RUS
113/769/0-3) and RFBR (under grant \# 08-02-91969). Work of EYK was
also supported by the Dynasty foundation. The authors are grateful
to T. Kinoshita and M. Nio for telling us details of their former
calculations and for confirming our result.

\end{document}